\documentclass[11pt]{article}
\usepackage{epsfig}
\usepackage{amsmath, amssymb}

\newcommand{\Eins}{{\mathbf 1}} 
\newcommand{\RR}{{\mathbb R}}
\newcommand{\ZZ}{{\mathbb Z}}
\newcommand{\HH}{{\cal H}} 
\newcommand{\UU}{{\cal U}} 
\newcommand{\QED}{\hspace*{\fill}$\square$}
\newcommand{\End}{\hbox{End}} 
\newcommand{\Hom}{\hbox{Hom}} 
\newcommand{\id}{\hbox{id}} 
\newcommand{\Tr}{\hbox{Tr\,}} 
\newcommand{\Ad}{\hbox{Ad}} 
\newcommand{\ind}{^{\rm ind}_+} 
\newcommand{\dual}{^{\rm dual}} 
\newcommand{\inv}{^{-1}} 
\newcommand{\rest}{\!\restriction}

\begin{document}

\title{On Local Boundary CFT  \\ and Non-Local CFT on the
  Boundary\footnote{Talk presented at ``Rigorous quantum field
    theory'', symposium in honor of J. Bros, Paris, July 2004.} }
\author{Karl-Henning Rehren \\ Institut f\"ur Theoretische Physik,
  Universit\"at G\"ottingen, \\ 37077 G\"ottingen, Germany, \\
  \texttt{rehren@theorie.physik.uni-goe.de}}  

\maketitle

\begin{abstract} The holographic relation between local boundary
  conformal quantum field theories (BCFT) and their non-local boundary
  restrictions is reviewed, and non-vacuum BCFT's, whose existence was
  conjectured previously, are constructed. (Based on joint work
  \cite{LR2} with R. Longo)
\end{abstract}

{\small PACS 2003: 03.70.+k.
}

\section{Introduction}
\setcounter{equation}{0} 
This contribution highlights some aspects of a recent analysis of 
relativistic conformal QFT in the presence of a boundary \cite{LR2}. 
The main result in \cite{LR2} is that the local observables of a
conformal field theory on  the half-space $x>0$ of two-dimensional
Minkowski space-time (``local boundary conformal QFT'' or BCFT for
short) can be reconstructed from their restrictions to the boundary,
which define a non-local chiral conformal QFT (``non-chiral local
fields arise from non-local chiral fields''), and vice versa. This
fact may be regarded as a ``holographic'' relation between quantum
field theories in different spacetime dimensions.  

A more detailed statement of the above is the following: A local
conformal QFT on the half-space $M_+=\{(t,x): x>0\}$ contains a
subalgebra of chiral fields, which may be naturally identified with
its restriction to the boundary, i.e., a local CFT on the real
line. Restricting the full BCFT to the boundary, one obtains 
a chiral CFT which is non-local, but relatively local with respect to
its chiral subtheory. Conversely, every chiral CFT which is relatively
local with respect to a given local chiral subtheory, induces a local
BCFT on $M_+$. Under some (rather natural) technical
assumptions, if one restricts a BCFT to the boundary and induces
another BCFT from the restriction, then the latter is the maximal
local extension of the original theory on the same Hilbert space. In
particular, every BCFT is a subtheory of one arising by induction from
some non-local chiral CFT.

This fact gives fresh motivation to study non-local chiral quantum
field theories. E.g., to regard the non-local chiral theories on the
boundary as the primary objects, opens a new route to classification
of boundary CFT. Structure results on non-local chiral QFT have 
immediate consequences for the induced BCFT's (and their sub-theories). 

While the statement of the main result amounts to the very simple
identity (2.8) below, its proof is rather involved. It turns out
advantageous to swap between the two-dimensional (BCFT) and the
one-dimensional (chiral CFT) point of view. This ``holographic''
attitude is particularly powerful in combination with Modular Theory
(\cite{B,T}, see below). In order to emphasize this situation, we
reorganize in these notes the line of argument and focus the
attention on the holographic and modular aspects in the interplay
between chiral CFT and boundary CFT, rather than doubling the approach
in \cite{LR2}.    

The Modular Theory of von Neumann algebras \cite{T} (briefly reviewed
in the appendix) is
most naturally tied to the algebraic approach to QFT \cite{H}. Many
results of great generality about local QFT have been obtained, when
this theory is applied to the algebras of bounded local observables in
suitable space-time regions in conjunction with the vacuum vector
\cite{B}. The most prominent is the Bisognano-Wichmann (BW) property
\cite{BW} in local QFT, which states that the modular group
$\Delta^{it}$ associated with the von Neumann algebra of observables
in a wedge region and the vacuum vector, coincides with the unitary
group of Lorentz boosts which preserve the wedge, hence the boosts are
of modular origin. In fact the entire Poincar\'e group (including
positivity of the energy spectrum \cite{W}) can be constructed from
modular groups of local algebras \cite{KW}. These methods are
particularly powerful in local chiral QFT where the conformal group
was found to be of modular origin \cite{GLW}.  

For non-local chiral theories, some nontrivial results concerning
the modular group of interval algebras have been previously obtained
\cite{DLR} for $\ZZ_2$-graded local (i.e., fermionic) theories,
including the Bisognano-Wichmann property. We show that these results
generalize to non-local chiral CFT's, whenever these are relatively
local with respect to a subtheory (which is automatically local) which
is contained with finite index; in particular, this is true whenever
the subtheory is the fixed point subalgebra under the action of a
finite internal symmetry group.   

In Sect.\ 4 of this contribution, we construct positive-energy
representations of chiral extensions associated with ``nimreps'' 
(non-negative integer matrix representations) of the fusion rules of
the underlying chiral observables. By the holographic relation, these
theories give rise to local ``non-vacuum'' boundary CFT's which were
conjectured to exist in \cite{LR2}, while special cases were
constructed explicitly in terms of chiral exchange fields \cite{P}. 
Their general construction strengthens the parallelism between local
BCFT and Euclidean BCFT in Statistical Mechanics, where analogous
theories are widely considered for the remarkable properties of
their partition functions $\Tr\exp-\beta L_0$ and their relation to
matrix elements of a two-dimensional heat kernel 
$\exp-(2\pi^2/\beta)(L^+_0+L^-_0)$ at inverse temperature, between
pairs of states from a finite family of distinguished improper
``boundary states'' \cite{Z}.

\section{Algebraic boundary conformal QFT}
\setcounter{equation}{0} 
We denote by $M_+$ the half-space $M_+=\{(t,x): x>0\}$, and by
$I\times J$ the set $\{(t,x): t+x \in I,\; t-x \in J
\}$. If $I$ and $J$ are open intervals in $\RR$, then $O=I\times J$ is a
double-cone in Minkowski spacetime, and $O\subset M_+$ iff $I>J$ elementwise.

A BCFT is a local quantum field theory on $M_+$, covariant under the 
subgroup of M\"ob $\times$ M\"ob (acting on $t+x$ and $t-x$ separately)
which preserves the boundary. This is the diagonal subgroup, and will
be identified with the M\"obius group itself.   

A BCFT contains certain chiral observables, e.g., the stress-energy
tensor or conserved currents. Due to a boundary condition at
$x=0$, the left and right chiral fields of the BCFT coalesce
and are identified with a local chiral field defined on the boundary $\RR$. 
Let $A(I)$ denote the von Neumann algebras generated by the latter
chiral observables smeared over an interval $I\subset \RR$. The
inclusion-preserving assignment $I\mapsto A(I)$ is called a 
{\em chiral net}. The chiral net $A$ is local and M\"obius covariant (and in
fact extends to a net on the circle, into which $\RR$ is embedded via a
Cayley transformation). We shall henceforth assume that the 
chiral net $A$ is {\em completely rational} \cite{KLM}, i.e., it is
split ($A(I)\vee A(J)\simeq A(I)\otimes A(J)$ if $I$ and $J$ are
disjoint without common boundary points) and the four-interval
subfactor has finite index, implying strong additivity \cite{LX}. 

The von Neumann algebras generated by the chiral observables of the
BCFT smeared in a double-cone $O=I\times J \subset M_+$ are  
$$A_+(O)= A(I) \vee A(J).\eqno(2.1)$$
This formula expresses the fact that smearing the BCFT chiral fields
over $O=I\times J$ is the same as smearing the chiral fields on $\RR$
over $I \cup J$. 

In general, a BCFT will contain local observables beyond the chiral
ones: A BCFT is a M\"obius covariant local net $O\mapsto B_+(O)$ 
where $O\subset M_+$ and $B_+(O)$ contain $A_+(O)$. These von Neumann
algebras act on a Hilbert space which is in general reducible as a
representation of $A_+$. We call this representation $\pi$, thus 
$$\pi(A_+(O))\subset B_+(O).\eqno(2.2)$$
The representation $\pi$ of $A_+$ is at the same time a representation
of the chiral net $I\mapsto A(I)$, and we require it to be a covariant 
positive-energy representation. We also assume that the algebra
generated by $\pi(A)$ together with any single local algebra $B_+(O)$
acts irreducibly on the Hilbert space of $B_+$ (``joint irreducibility''), 
expressing the physical property that the stress-energy tensor
(contained in $A$) locally generates the translations of the BCFT, and
hence together with a single local algebra $B_+(O)$ generates the entire net. 
 
Our aim is to understand the structure of the extension (2.2). 
It is the most remarkable conclusion of our work that this is possible
in terms of (local or non-local) {\em chiral} nets which contain the
chiral observables.  

A {\em chiral extension} of $A$ is a M\"obius covariant net of
inclusions 
$$\pi(A(I))\subset B(I)\eqno(2.3)$$
in the vacuum representation of $B$, such that $B$ and $A$ are
relatively local, i.e., $B(I_1)$ commute with $\pi(A(I_2))$
whenever $I_i$ are disjoint. $B$ contains $A$ irreducibly if the
inclusions (2.3) are irreducible. In this case, $B(I)$ are factors and
(2.3) are subfactors. 

The passage between chiral and boundary CFT makes use of two basic
operations: ``restriction'' and ``induction''. Restriction associates
a chiral net (over the intervals $I\subset \RR$) with a given BCFT net
(over the double-cones $O\subset M_+$), and induction associates a
BCFT net with a given chiral net.   

Every interval $I\subset\RR$ defines a left wedge $W_L(I)=\{(t,x) \in
M_+: t+x \in I, t-x \in I\}$, and a right wedge $W_R(I):=W_L(I)'$ (the
interior of the causal complement within $M_+$). 

Let a BCFT $B_+$ be given. For either wedges, let
$B_+(W):= \bigvee_{O\subset W} B_+(O)$. The {\em boundary net} 
$I\mapsto \partial B_+(I)$ is defined by {\em restriction} of $B_+$ to
the left wedges, 
$$\partial B_+(I):= B_+(W_L(I)).\eqno(2.4)$$
For the time being, we also introduce the net $I\mapsto \partial'B_+(I)$ 
defined by 
$$\partial'B_+(I):= B_+(W_R(I))'\eqno(2.5)$$
but we shall later see that this net coincides with $\partial B_+$.
Both chiral nets $B=\partial B_+$ or
$=\partial'B_+$ are chiral extensions of $A$; joint irreducibility of
$B_+$ implies irreducibility of the chiral extensions. But although
$B_+$ is local, $\partial B_+$ and $\partial' B_+$ are in general
non-local nets. 

Conversely, let a chiral extension $\pi(A) \subset B$ be given. 
For $O=I\times J\subset M_+$ and $J=(a,b) < I=(c,d)$, let $K$ and $L$
denote the intervals $K=(b,c)$ and $L=(a,d)$. The {\em induced net}
$O\mapsto B\ind(O)$ is defined by  
$$B\ind(O):=B(K)'\cap B(L). \eqno(2.6)$$
Even if $B$ may be non-local, $B\ind$ is local. If $B$ contains $A$
irreducibly, then $B\ind$ satisfies joint irreducibility. 
It follows that $B\ind$ is a BCFT net. 

One has the following relations between the two constructions. 
\vskip1.0mm 
{\bf Theorem:} Let $B$ be an irreducible local extension of $A$, and
let $B_+$ be a BCFT net. Then \\[0.4mm]
{\bf (T1)} $\partial' B_+ = \partial B_+$. \\[0.4mm]
{\bf (T2)} $\partial (B\ind) = B$. \\[0.4mm]
{\bf (T3)} $(\partial B_+)\ind = (B_+)\dual$.
\vskip1.0mm
The dual net in (T3) is defined standardly: since $O'$ is the
union of a left wedge $W_L$ and a right wedge $W_R$, let
$B_+(O'):=B_+(W_L)\vee B_+(W_R)$. Then 
$$(B_+)\dual(O):= B_+(O')'.\eqno(2.7)$$
The dual of a local net extends
the local net but needs not to be local itself; if it 
is local, then it equals its own dual, i.e., it is self-dual. 

Let us first discuss the far-reaching implications of this theorem.
\vskip1.0mm
{\bf Corollary:} 
{\bf (C1)} Every BCFT satisfies {\em wedge duality}, i.e.,
$B_+(W')=B_+(W)'$ for any wedge $W$. \\[0.4mm]
{\bf (C2)} Every irreducible chiral extension arises by restriction of some
BCFT. \\[0.4mm]
{\bf (C3)} Every self-dual BCFT net arises by induction from some irreducible
chiral extension. \\[0.4mm]
{\bf (C4)} The dual net of a BCFT is local, hence self-dual. \\[0.4mm]
{\bf (C5)} Every induced BCFT net is self-dual. \\[0.4mm]
{\bf (C6)} The maps $B\mapsto B\ind$ (induction) and $B_+\mapsto \partial
B_+$ (restriction) give a bijection between self-dual BCFT nets and
irreducible chiral extensions. 
\vskip1.0mm
(C1) is equivalent to (T1) by the definitions.
(C2) is obvious from (T2), and (C3) from (T3). (C4) follows from (T3)
because every induced net is local. (T2) and (T3) together imply (C5)
because $(B\ind)\dual=(\partial B\ind)\ind = B\ind$, as well as (C6)
because $\partial (B\ind) = B$ and $(\partial B_+)\ind = (B_+)\dual=B_+$. 
\QED

In particular, for self-dual BCFT nets, one has 
$$B_+(O)=B(K)'\cap B(L)\eqno(2.8)$$
($B=\partial B_+$ the corresponding boundary net, and $K$ and $L$ as
in (2.6) before). One may regard the identity (2.8) as a
``holographic relation'' between a two-dimensional CFT and a
one-dimensional (chiral) CFT: The inclusion (2.2) is completely
determined by the chiral extension (2.3), see Sect.\ 3 below. 

By (C6), classification of self-dual BCFT's is equivalent to
classification of irreducible chiral extensions. This is a finite
problem, because completely rational nets possess only finitely many
irreducible chiral extensions. E.g., for Virasoro nets with $c<1$ and
for $SU(2)$ chiral current algebras, complete classifications of local
chiral extensions have been obtained in \cite{KL}, and non-local ones
can be classified along the same lines. 

A general BCFT $B_+$ is intermediate between $\pi(A_+)$
and $(B_+)\dual$. Because one can show that $\pi(A_+(O))\subset
(B_+)\dual(O)$ given by (2.8) has finite index, the number of
intermediate nets is also finite. 

We disentangle the proof of the theorem by a series of lemmas, which
are of more or less interest of their own.
\vskip1.0mm
{\bf Lemmas:} (All nets and extensions here are assumed to be M\"obius
covariant. As before, chiral extensions are relatively
local. Otherwise, locality of the nets is stated explicitly if
assumed, and so is complete rationality.) \\[0.4mm]
{\bf (L1)} If a local chiral net $A$ is completely
rational, then every irreducible chiral extension of $A$ has finite
index.\\[0.4mm]
{\bf (L2)} If $B$ extends $A$ with finite index and $A$ is split, then
$B$ is split. (The split property for non-local nets is $B(K)\vee
B(L)' \simeq B(K)\otimes B(L)'$ if $K$ and $L$ are disjoint intervals
without common boundary points.)\\[0.4mm]
{\bf (L3)} If a chiral net is split, then 
$\bigcap_{K\subset L}[B(K)\vee B(L)'] = \left[\bigcap_{K\subset L}B(K)\right]
\vee B(L)' = B(L)'$.\\[0.4mm]
{\bf (L4)} Every chiral net $B$ has a maximal relatively local net
$C\subset B$. $C$ is unique, hence covariant and invariant under the
gauge group of $B$. Clearly, $C$ is local, and if $A\subset B$ is
relatively local then $A\subset C\subset B$.\\[0.4mm]
{\bf (L5)} If $C\subset B$ is invariant under the gauge group of $B$,
then there are local vacuum-preserving conditional expectations 
$B(I)\to C(I)$. \\[0.4mm]
{\bf (L6)} If $A\subset C$ and $C$ is local, then there are local
vacuum-preserving conditional expectations $C(I)\to A(I)$. \\[0.4mm]
{\bf (L7)} The local conditional expectations $B(I)\to A(I)$ of a
chiral extension (existing according to (L4--L6)) are consistent,
i.e., the expectations for two intervals coincide on the
intersection of the corresponding local algebras, and are implemented
by the projection on the subspace $\overline {A\Omega}$.  \\[0.4mm]
{\bf (L8)} If $A\subset B$ has finite index and there are local
vacuum-preserving conditional expectations $B(I)\to A(I)$,
then $B$ has the BW property if $A$ does. \\[0.4mm]
{\bf (L9)} If a chiral extension $B$ of $A$ has finite index, then $B$
has the BW property.
\vskip1.0mm
(L1) \cite{ILP} and (L2) \cite{LR2} employ subfactor theory techniques
(the generation of the larger algebra by the smaller and certain
isometries). Proving (L3) and (L4) is elementary \cite{LR2}. 
(L5--L9) invoke Modular Theory.

We focus on the steps involving Modular Theory (referring to the
appendix for its results used here). (L5) and (L6) follow  
with Takesaki's Theorem about conditional expectations and modular
stability of subalgebras as well as the characterization of the
departure from the BW property in non-local theories, based on
Borchers' modular commutation relations: in (L5), $C(I)$ is invariant
under the modular group of $(B(I),\Omega)$ by (A.5), and in (L6), $C$,
being local, satisfies the BW property, hence $A(I)$ is invariant
under the modular group of $(C(I),\Omega)$. So in both cases (i)
$\Rightarrow$ (ii) in Takesaki's Theorem (see the appendix) proves the
claim. (L7) uses the restriction and implementation properties in
Takesaki's Theorem: since $A$ is local, the cocycle for $A$ is
trivial, hence the cocycle $z(t)$ for $B$ is trivial on the cyclic
subspace of $A$, hence the modular group of $B(I)$ acts trivially on
$A(I)$. (L7) follows because the projection on $\overline
{A(I)\Omega}$ is independent of $I$ (Reeh-Schlieder theorem). By the
same argument, it follows in (L8) that the fixed point subalgebra
$B(I)^z$ of $B(I)$ under the one-parameter group $\Ad_{z(t)}$ of
automorphisms contains $A(I)$, hence the index of the fixed point
subalgebra is finite. Because $\RR$ has no nontrivial finite
quotients, we must have $B(I)^z=B(I)$, hence $B(I)$ commutes with
$z(t)$. $\Omega$ being cyclic for $B(I)$, this implies that $z(t)$ is
trivial, hence (L8). Combining (L4--L8), gives (L9) because $A$, being
local, has the BW property.  \QED

Now the theorem is proven easily. (L1--L3) mean that 
$\bigvee_{K\subset L} B\ind(O)=B(L)$, which is the statement of
(T2). By (L1) and (L9) we conclude that $\partial B_+(I)$ and
$\partial'B_+(I)$ have the same modular group (namely the dilations of
$I$). Because $\partial B_+(I) \subset \partial' B_+(I)$ and $\Omega$
is cyclic for both algebras, the implementation property in Takesaki's
Theorem implies equality. This is (T1). (T3) is then obvious by
writing (2.7) as $\partial B_+(K) \cap \partial'B_+(L)$. \QED

Along the way, we proved the following proposition
(assembling results on chiral extensions) and its corollary
(assembling the implications for BCFT):
\vskip1.0mm
{\bf Proposition:} 
{\bf (P1)} Every chiral extension has a consistent family of
vacuum-preserving conditional expectations ${\cal E}^I:B(I)\to A(I)$. \\[0.4mm]
{\bf (P2)} If a chiral net $B$ is a chiral extension with finite index
of a local net $A$, then $B$ satisfies the BW
property. In particular, this is the case for all irreducible chiral
extensions of completely rational nets. \\[0.4mm]
{\bf (P3)} The split property is upward hereditary for chiral
extensions of finite index. 
\vskip1.0mm
(P1) is the combination of (L4-L7). (P2) is (L8), using the fact that
a local chiral net has the BW property. (P3) is (L2). \QED

The existence of a consistent family of vacuum-preserving conditional
expectations has been a crucial assumption, expressing a generalized symmetry
principle, in the general structure theory for chiral extensions
\cite{LR}. We see from (P1--P3) that it is automatic for boundary
nets of a BCFT which are irreducible extensions of a completely
rational chiral net. Thus, subfactor methods as developped in
\cite{LR} may be applied.  
\vskip1.0mm
{\bf Corollary:} 
{\bf (C7)} Every BCFT satisfies the split property for wedges. \\[0.4mm]
{\bf (C8)} The index of $\pi(A_+(O))\subset B_+(O)$ is universal for
self-dual BCFT nets $B_+$ (i.e., it depends only on the chiral
observables $A$). \\[0.4mm]
{\bf (C9)} A self-dual BCFT is strongly additive, i.e., if $O_i$ are two
spacelike separated double-cones which touch each other in one
point and $O=I\times J$ is the double-cone spanned by $O_1$ and $O_2$,
then $B_+(O_1)\vee B_+(O_2)=B_+(O)$. \\[0.4mm] 
{\bf (C10)} A self-dual BCFT satisfies Haag duality for finite unions
of spacelike separated double-cones. \\[0.4mm]
{\bf (C11)} A self-dual BCFT net has no DHR sectors. 
\vskip1.0mm
(C7) follows for induced BCFT nets by the definition because $B$ is
split by (P3) because $A$ is completely rational. Then (C7) is true
for any BCFT because $B_+$ is contained in $(B_+)\dual$, which is an
induced net. For (C8), let $B=\partial B_+$ denote the boundary net,
and consider the chain of inclusions
$$\pi(A(K))\vee \pi(A(L'))\subset \pi(A(K))\vee B(L)' \subset 
B(K) \vee B(L)' \subset \pi(A(I))'\cap \pi(A(J))'. $$
By the split property (P3) for $B$, the first two inclusions both have
index $[B:A]$, and by the general theory in \cite{KLM}, the total
index equals $d(\pi)^2\mu_A$ where $d(\pi)=[B:A]$ and $\mu_A$ is the
``dimension'' of the DHR superselection category of $A$, i.e., the sum
of the squares of the dimensions of all irreducible DHR sectors of
$A$. Hence, the index of the last inclusions is $\mu_A$, and by 
passing to the commutants, this is the index of the subfactor in (C8). 

(C8) implies (C9) by standard subfactor methods \cite{L}, relying only
on the finiteness of the index in (C8). By similar methods as for
(C8), the index of $B_+(E)\subset B_+(E')'$ when
$E$ is a finite union of spacelike separated double-cones, is computed
\cite{LR2} to be a power of $[(B_+)\dual:B_+]$, hence it is
1 if $B_+$ is self-dual, implying (C10). (C10) implies (C11) by standard
arguments. \QED

\section{The charge structure of BCFT fields}
\setcounter{equation}{0} 

The holographic formula (2.8) allows for an explicit computation of
the local algebras $B_+(O)$ of a self-dual boundary CFT in terms of
the chiral extension $A\subset B$. 
It is found \cite{LR2} that $B_+(O)$ is generated by $A_+(O)$ along
with a finite system of charged BCFT operators, carrying a product
of chiral charges which is bi-localized in $I$ and in $J$ if 
$O=I\times J$.  

The notion of chiral charge here refers to the irreducible superselection
sectors (generalized charges) of the theory $A$. The bi-localized charge
structure expresses itself in the commutation relations with the local
fields on the boundary: A charged operator $\psi\in B_+(O)$ is an
intertwiner for $\sigma\circ\bar\tau$ where $\sigma$ and $\bar\tau$
are DHR endomorphisms of $A$ localized in $I$ and in $J$,
respectively. Hence $\psi$ carries a charge $[\sigma]$
localized in $I$ and a charge $[\bar\tau]$ localized in $J$. 

The precise combinatorial structure (including the pairing of
chiral charges) of these local charged intertwiners in terms of
non-local ``chiral vertex operators'' can be determined by solving a
certain eigenvalue problem, depending on the chiral extension
$A\subset B$, within the DHR superselection category of $A$.  

The admissible pairing of chiral charges is described by a modular
invariant matrix $Z_{\sigma,\tau}$. It is also proven in \cite{LR2}
that the induced BCFT $B\ind$ is locally isomorphic to
a local two-dimensional CFT $B_2$ on the entire Minkowski spacetime
(whose Hilbert space is given by the same modular invariant matrix
$Z$), previously obtained by a ``generalized quantum double
construction'' \cite{CTPS}. There is thus a common (sub)net, defined
sufficiently far away from the boundary, of which $B_2$ and $B\ind$
may be regarded as different quotients (representations) distinguished
by the absence or presence of the boundary.

\section{Nimreps and non-vacuum BCFT}
\setcounter{equation}{0} 
This section closes a gap which was left open in \cite{LR2}. 
We canonically associate with a given non-local chiral extension
$\pi(A)\subset B$ a family of chiral extensions $B_b$ along with a
family of positive-energy representations $\pi^{ab}$ of $B_b$ where
the labels $a$ and $b$ (``boundary conditions'') run over the same
finite set. The holographic construction yields a corresponding family
of BCFT's $B_{+,ab}$.
 
The multiplicities $n_{ab}^\rho$ of the irreducible DHR sectors
$[\rho]$ of the chiral observables $A$ within $\pi^{ab}$ form a
normalized ``nimrep'' (non-negative integer matrix representation) of
the fusion rules of the DHR sectors of $A$, 
$$ n^\sigma\cdot n^\rho = \textstyle\sum_\tau N^\tau_{\sigma\rho}\; n^\tau  
\qquad\hbox{such that}\qquad n^0_{ab}=\delta_{ab}.\eqno(4.1)$$
In particular, unless
$a=b$, these representations do not contain the vacuum sector. 
The property (4.1) implies that the partition functions $\Tr\exp-\beta
L_0$ of $B_{+,ab}$ coincide with matrix elements of the heat kernel 
$\exp-(2\pi^2/\beta)(L^+_0+L^-_0)$ of the associated Minkowski theory
$B_2$ between pairs of states $\vert a\rangle$ and $\vert b\rangle$
from a finite family of distinguished improper ``boundary states''
\cite{Z}.       

Our construction starts by ``acting with the DHR sectors of $A$'' on
$\pi(A)\subset B$ as in \cite{BE,LR2}: 
Let $\iota$ denote the inclusion homomorphism $\iota:A(I)\to B(I)$ for
a fixed interval $I$. Then consider homomorphisms
$\iota\circ\sigma:A(I)\to B(I)$ where $\sigma$ 
runs over the DHR endomorphisms of $A$ localized in $I$, and let $X$ be
the (finite) set of all equivalence classes $[a]$ of 
irreducible subhomomorphisms\footnote{A homomorphism $\beta:N\to M$ is
  a subhomomorphism ($\beta\prec\alpha$) of $\alpha:N\to M$, if the
  intertwiner space $\Hom(\beta,\alpha):=\{t\in M:
  t\beta(n)=\alpha(n)t\;\forall \,n\in N\}$ contains an isometry,
  $t^*t=1$; hence $\beta(n)=t^*\alpha(n)t$. }
$a\prec \iota\circ\sigma$ as $\sigma$ varies. For every pair $a,b$ in
$X$ and $\bar a:B(I)\to A(I)$ a conjugate\footnote{A homomorphism
  $\bar\alpha:M\to N$ is conjugate to $\alpha:N\to M$, if
  $\id_N\prec\bar\alpha\circ\alpha$ and
  $\id_M\prec\alpha\circ\bar\alpha$. Since $A$ is assumed to be
  completely rational, all homomorphisms in the sequel decompose into
  finitely many irreducibles, and possess conjugates.} homomorphism of
$a$, the product $\bar a\circ b:A(I)\to A(I)$ is (the local
restriction of) a DHR endomorphism localized in $I$. It follows that
the numbers\footnote{Strong additivity of $A$ guarantees
  equivalence between local and global intertwiners, hence the notions
  of equivalence and subendomorphisms are the same for DHR
  endomorphisms and for their local restrictions.}   
$$n^\rho_{ab} = \dim\Hom(\rho,\bar a\circ b) \equiv
\dim\Hom(a\circ\rho,b) \eqno(4.2)$$
form a nimrep as above, including the normalization. 

The construction of the family of chiral CFT's and hence BCFT's associated
with this nimrep now relies on the following equivalence.
\vskip1.0mm
{\bf Theorem} \cite{LR}: There is a 1:1 correspondence (up to
unitary resp.\ algebraic equivalences) between \\[0.4mm]
{\bf (i)} irreducible covariant chiral extensions $\pi(A)\subset B$, defined
on the vacuum Hilbert space of $B$. \\[0.4mm]
{\bf (ii)} irreducible subfactors $A_1\subset A(I_0)$ (for any fixed
interval $I_0$) whose canonical endomorphism\footnote{If $N\subset M$
  with inclusion homomorphism $\iota$ and conjugate $\bar\iota$, then
  $\bar\iota\circ\iota\in\End(N)$ is the canonical endomorphism, and
  $\iota\circ\bar\iota\in\End(M)$ the dual canonical endomorphism.} is
(the local restriction of) a DHR endomorphism $\theta$ of $A$
localized in $I_0$.  
\vskip1.0mm
More precisely \cite[Cor.\ 3.3 and Prop.\ 3.4]{LR}: a chiral extension
gives an inclusion homomorphism $\iota:A\to B$ such that
$\pi=\pi^0\circ \iota$, where $\pi^0$ is the defining vacuum
representation of $B$. For any interval $I_0$, one can construct a
conjugate homomorphism $\bar\iota:B\to A$ such that
$\bar\iota\rest_{B(I)}:B(I)\to A(I)$ is conjugate to
$\iota\rest_{A(I)}:A(I)\to B(I)$ whenever $I\supset I_0$, and
$\theta=\bar\iota\circ\iota$ is a DHR endomorphism of $A$ 
localized in $I_0$. Then $\pi^0$ is unitarily equivalent to
$\pi_0\circ\bar\iota$, where $\pi_0$ is the defining vacuum
representation of $A$, and consequently $\pi$ is unitarily equivalent
to $\pi_0\circ\theta$. This unitary equivalence turns the 
covariance $\UU$ of $B$ into the covariance\footnote{The
  covariances here are the unitary representations of the M\"obius
  group such that $\Ad_{\UU(g)} B(I) = B(gI)$ and
  $\Ad_{\UU_\theta(g)}\circ \pi_0\circ\theta(A(I))  =
  \pi_0\circ\theta(A(gI))$, respectively.} 
$\UU_\theta$ of $A$ in the DHR representation $\pi_0\circ\theta$.  

The 1:1 correspondence in the Theorem is given by the subfactor 
$$A_1 := \bar\iota(B(I_0)) \subset A(I_0).\eqno(4.3)$$
From the subfactor $A_1\subset A(I_0)$, one can reconstruct
$A(I_0)\subset B(I_0)$ by the Jones basic construction \cite{J}, along
with a conditional expectation which extends the vacuum state on
$A(I_0)$ to a state on $B(I_0)$. The construction of the entire net
$B$ is then achieved with the help of charge transporters
or using the M\"obius covariance \cite[Thm.\ 4.9]{LR}. The vacuum
state extends to the net, and its GNS representation is the vacuum
representation of $B$.

Now, let a chiral extension $\pi(A)\subset B$ be given. Fix an
interval $I_0$ and let $\iota:A(I_0)\to B(I_0)$ denote the local inclusion
homomorphism. Then $\theta=\bar\iota\circ\iota$ is (the local restriction
of) a DHR endomorphism localized in $I$. Let $\sigma$ be a DHR
endomorphism localized in $I_0$ and $a\prec\iota\circ\sigma$ an irreducible
sub-homomorphism as before. Then the subfactor $A_{1,a}:=\bar
a(B(I_0))\subset A(I_0)$ has canonical endomorphism $\theta_a=\bar a\circ a$
which is (the local restriction of) a DHR endomorphism because 
$\theta=\bar\iota\circ\iota$ is (the local restriction of) a DHR
endomorphism and $\bar a\circ a\prec\bar\sigma\circ\theta\circ\sigma$. By 
(iii)$\Rightarrow$(i), the subfactor $A_{1,a}\subset A(I)$ defines
a chiral extension $\pi_a(A)\subset B_a$. 
 
This construction results in a finite family of chiral extensions in
their vacuum representations (called the ``DHR orbit'' in \cite{LR2};
note that inequivalent $a\in X$ may result in equivalent extensions,
e.g., if $a_1=a_2\circ\tau$ where $\tau$ is a DHR automorphism). 

Each element of this family generates the whole family (and the same
nimrep) by the same construction \cite{BE}: namely, starting from
$A(I_0) \subset B(I_0)$, let $c\in X$, and construct $A(I_0)\subset
B_c(I_0)$ as before with inclusion homomorphism $\iota_c:A(I_0)\to
B_c(I_0)$ and dual canonical endomorphism
$\theta_c=\bar\iota_c\circ\iota_c=\bar c\circ c$.  
Then because $\bar c(B(I_0))=A_{1,c}=\bar\iota_c(B_c(I_0))$, the mapping
$\varphi_c:=\bar c\inv\circ\bar\iota_c:B_c(I_0)\to B(I_0)$ is well
defined, and is an algebraic isomorphism between $B_c(I_0)$ and
$B(I_0)$. Using $A(I_0)\subset B_c(I_0)$ as the starting point instead,
denote by $X_c$ the set of equivalence classes of irreducible
subhomomorphisms $a_c:A(I_0)\to B_c(I_0)$ of $\iota_c\circ\sigma$ for DHR
endomorphism $\sigma$ localized in $I_0$. Then  
$\varphi_c\circ a_c:A(I_0)\to B(I_0)$ is contained in $\bar
c\inv\circ\bar\iota_c\circ\bar\iota_c\circ\sigma = \bar c\inv\circ\bar c\circ 
c\circ\sigma  = c\circ\sigma \prec \iota\circ\sigma'\circ\sigma$
because $c$ is contained in $\iota\circ\sigma'$ for some $\sigma'$. 
Hence $\varphi_c\circ a_c$ belongs to $X$. Conversely, if
$a\prec\iota\circ\sigma$ belongs to $X$, then 
$a\prec c\circ\bar\sigma'\circ\sigma$ for some $\sigma'$, hence 
$\varphi_c\inv\circ a\prec
\bar\iota_c\inv\circ\bar c\circ c \circ \bar\sigma'\circ\sigma =
\iota_c\circ\bar\sigma' \circ \sigma$
belongs to $X_c$. 

This means that $\varphi_c$ provides by left composition a bijection
between the irreducible subhomomorphisms of
$\iota\circ\sigma:A(I_0)\to B(I_0)$ as
$\sigma$ varies, and those of $\iota_c\circ\sigma:A(I_0)\to B_c(I_0)$
as $\sigma$ varies. Moreover, $\bar a_c(B_c(I_0)) = \bar
a_c\circ\varphi_c\inv(B(I_0)) = \bar a(B(I_0))$ 
give the same subfactors $A_{1,a}\subset A(I_0)$, and hence the same 
family of chiral extensions. Obviously, the bijection preserves the
nimrep $n^\rho_{ab} = \dim\Hom(\rho,\bar a\circ b)$.

We now turn to positive-energy representations of chiral extensions,
using $\alpha$-induction \cite{LR}. $\alpha$-induction
covariantly extends a DHR endomorphism $\sigma$ of $A$ to an endomorphism
of its extension $B$, and hence to a covariant induced representation
of $B$. The latter is unitarily
equivalent to $\pi_0\circ\sigma\circ\bar\iota$ \cite{LR}. This
representation of $B$ restricts to the DHR representation
$\pi_0\circ\sigma\circ\bar\iota\circ\iota = 
\pi_0\circ\sigma\circ\theta$ of $A$. Since the
covariance of $B$ in the representation $\pi_0\circ\bar\iota$ is
$\UU_\theta$, the covariance in induced representation
$\pi_0\circ\sigma\circ\bar\iota$ is
$\UU_\sigma\sigma(\UU_0^*\UU_\theta)=\UU_{\sigma\circ\theta}$
\cite{FRS}. The induced representation is in general reducible. 

Let $a,b\in X$. Then $\bar a\circ b$ is the local restriction of a
DHR endomorphism $\theta_{ab}$ with multiplicities $n^\rho_{ab}$. 
Moreover, there is a DHR endomorphism $\sigma$ of $A$ localized in
$I_0$ and an isometric local intertwiner $t\in\Hom(\bar 
a,\sigma\circ\bar b)\subset A(I_0)$. We want to show that $t$ reduces 
the induced representation $\pi_0\circ\sigma\circ\bar\iota_b$ of
$B_b$. Namely, $t$ is an intertwiner between the local restrictions of
the DHR endomorphisms $\theta_{ab}=\bar a\circ b$ and
$\sigma\circ\theta_b = \sigma\circ\bar b\circ b$, and 
hence it is also a global intertwiner. The projection $tt^*$ thus 
commutes with $\sigma\circ\bar b(B(I_0)) =
\sigma\circ\bar\iota_b(B_b(I_0))$ (i.e., locally) and with 
$\sigma\circ\bar\iota_b\circ\iota_b(A)$ (globally). Since $B_b(I_0)$
and $A$ generate the entire net $B_b$, $tt^*$ commutes globally with
$\sigma\circ\bar\iota_b(B_b)$. 

Thus $tt^*$ defines a subrepresentation $\hat\pi^{ab} =
\pi_0\circ\Ad_{t^*}\circ\sigma\circ\bar\iota_b$ of the induced
representation of $B_b$, which restricts to the DHR representation 
$\pi_0\circ\Ad_t^*\circ\sigma\circ\theta_b = \pi_0\circ\theta_{ab}$ of 
$A$. Because $t\in\Hom(\theta_{ab},\sigma\circ\theta_b)$ also
intertwines the covariances of the DHR endomorphisms \cite{FRS},
$\hat\pi^{ab}$ is covariant with covariance $\UU_{\theta_{ab}}$.

By choosing a unitary operator $U$ to transport $\pi_0\circ\theta_{ab}$
to a representation $\pi_{ab}$ of $A$ on $\HH_{ab}= \bigoplus_\rho
n^\rho_{ab}\HH_\rho$, we obtain the desired covariant extensions 
$$\pi_{ab}(A(I)) \subset \pi^{ab}(B_b(I)) \eqno(4.4)$$
with
$$\pi^{ab} := \Ad_U\circ\pi_0\circ\Ad_{t^*}\circ\sigma\circ\bar\iota_b.
\eqno(4.5)$$
These theories are positive-energy representations (subrepresentations of
$\alpha$-induced representations) of the chiral extensions $B_b$,
which were previously defined in their vacuum representations. \QED

Note that in (4.5) we cannot write $\Ad_{t^*}\circ\sigma\circ\bar\iota_b = 
\Ad_{t^*}\circ\sigma\circ\bar b\circ\varphi_b = \bar a\circ\varphi_b$
because $\varphi_b$ is only defined on $B_b(I_0)$ while $\bar\iota_b$
extends to the net $B_b$. Nevertheless, $\pi^{ab}$ depends only on
$a$ and $b$, and not on the choice of $\sigma$ and $t\in\Hom(\bar
a,\sigma\circ\bar b)$. Namely, this is true for $\pi^{ab}(B_b(I_0)) =
\Ad_U\circ\pi_0\circ\bar a\circ\varphi_b(B_b(I_0)) =
\Ad_U\circ\pi_0\circ\bar a(B(I_0))$, and $\pi^{ab}(B_b(I))$ is
obtained by acting with the covariance of
$\Ad_U\circ\pi_0\circ\theta_{ab}$ on $\pi^{ab}(B_b(I_0))$. 

In the case $a=b$, we may choose $\sigma$ trivial and $t=1$. Thus
$\pi^{aa}(B_a)$ coincides with the chiral extension $B_a$ in its
vacuum representation. 
\vskip1.0mm
{\bf Corollary:} The same ``holographic construction'' as in (2.6):
$$ B_{+,ab}(O):= \pi^{ab}(B_b(K))' \cap \pi^{ab}(B_b(L)) \eqno(4.6) $$
for all $a,b\in X$ defines a family of covariant BCFT's containing
$\pi_{ab}(A_+)$ on the Hilbert spaces $\HH_{ab}$ with multiplicities
given by the nimrep (4.2). 
\vskip1.0mm
The corollary is obvious by the results of \cite{LR2} mentioned in Sect.\ 2. 
\QED

\appendix

\section{Modular Theory in QFT and in BCFT}
\setcounter{equation}{0} 
We assemble those aspects of Modular Theory, which are particularly
relevant for BCFT (and for QFT in general). We recommend also \cite{B}
and \cite{T}.

The fundamental result of Modular Theory is the following.
\vskip1.0mm
{\bf Tomita's Theorem} \cite[Chap.\ VI, Thm.\ 1.19]{T}: Let $M$ be a
von Neumann algebra on a Hilbert space $\HH$ with a cyclic and
separating vector $\Omega$. Then the anti-linear operator
$S:m\Omega\mapsto m^*\Omega$ is closable, and its closure has the
polar decomposition $\overline S = J\Delta^{\frac12}$ where
$J=J_{(M,\Omega)}$ is an anti-unitary involution and
$\Delta=\Delta_{(M,\Omega)}\geq 0$ is an invertible positive
self-adjoint operator such that $\Delta^{it}$ is a unitary
one-parameter group. These ``modular data'' have the properties \\[0.4mm]  
{\bf (i)} $\Delta J = J\Delta\inv$, $\Delta^{it} J = J \Delta^{it}$. \\[0.4mm]
{\bf (ii)} $\Delta^{it}$ implements a one-parameter group of automorphisms
$\sigma_t$ of $M$, i.e., 
$$\sigma_t(M)=M\qquad\hbox{where}\qquad \sigma_t(m):=\Delta^{it}m\Delta^{-it}.
\eqno(A.1)$$ 
{\bf (iii)} The conjugation $J$ maps $M$ onto its commutant, i.e.,
$$j(M)=M' \qquad\hbox{where} \qquad j(m):=JmJ.\eqno(A.2)$$ 
{\bf (iv)} The state $\omega=(\Omega,\cdot\;\Omega)$ is a KMS state of
inverse temperature $\beta=1$ on $M$ with respect to the
``dynamics'' given by the inverse modular automorphism group
$\sigma_{-t}$. The modular automorphism group is determined by this
property.  
\vskip1.0mm
The modular data contain highly nontrivial information about the
``position'' of a von Neumann algebra in its Hilbert space relative to the
distinguished state. In local QFT, this information is of dynamical
nature. The Bisognano-Wichmann theorem \cite{BW} states that, if
$\Omega$ is the vacuum vector and $M$ the von Neumann algebra of
observables localized in a wedge region, then the modular conjugation
$J$ is a CPT operator (which in asymptotically complete QFT is related
to the scattering matrix), and the modular group $\Delta^{it}$
coincides with the unitary group of Lorentz boosts which preserve the
wedge:   
$$\Delta^{it}_{(A(W),\Omega)}=U(\Lambda_W(-2\pi t)).\eqno(A.3)$$
By the statement (iv), the BW property (A.3) explains the Unruh effect,
that an accelerated observer whose ``dynamics'' is given by the boosts
in the Rindler wedge senses the vacuum state like a thermal state. 
Also the Hawking effect can be explained in this way,
because an observer of a black hole ``at rest'' in constant distance
from the surface, is in fact an accelerated observer. 

Since, as the wedges vary, the corresponding Lorentz boosts generate
the Poincar\'e group, the BW property implies that the unitary
representation of the Poincar\'e group is of modular origin. In a more
ambitious program \cite{KW}, one attempts to derive the Poincar\'e
group from the modular data of a finite set of wedge algebras, which
are required to be in a suitable ``relative modular position'' such as
to ensure the correct relations among their respective modular groups
in order to generate the Poincar\'e group.

In chiral theories, wedges are replaced by $\RR_+$ and the Lorentz
boosts by the dilations. By conformal covariance, this situation is
transported to any interval and the subgroup of M\"ob which preserves
the interval. Thus, the BW property is the statement that the modular
group of an interval algebra in the vacuum coincides with the unitary
representation of this subgroup. Since, as the interval varies, these
subgroups generate the M\"obius group, the BW property implies that
the conformal group is of modular origin.  

Of relevance for BCFT are mainly the following two profound theorems. 
\vskip1.0mm
{\bf Takesaki's Theorem} \cite[Chap.\ IX, Thm.\ 4.2]{T}: Let
$N\subset M$ be a pair of von Neumann algebras on a Hilbert space
$\HH$ and $\Omega\in \HH$ a vector which is cyclic and separating for
$M$. Then the following are equivalent: \\[0.4mm] 
{\bf (i)} $N$ is globally invariant under the modular automorphism group of
$(M,\Omega)$, i.e., $\sigma^{(M,\Omega)}_{t}(N)=N$.  \\[0.4mm]
{\bf (ii)} There exists a conditional expectation ${\cal E}:M\to N$ which
preserves the state $\omega=(\Omega,\cdot\;\Omega)$, i.e.,
$\omega\!\restriction_N({\cal E}(m)) = \omega(m)$.

In this case, ${\cal E}$ is implemented by the projection $E$ onto the
cyclic subspace ${\overline N\Omega}$, i.e., $EmE= {\cal E}(m)E$. 
Moreover, the modular data associated with $(M,\Omega)$ restrict on
the cyclic subspace to those associated with $(N,\Omega)$, i.e.,
$E\Delta_{(M,\Omega)}= \Delta_{(M,\Omega)}E = \Delta_{(N,\Omega)}$ and
$EJ_{(M,\Omega)}= J_{(M,\Omega)}E = J_{(N,\Omega)}$. In particular, if
$\Omega$ is cyclic also for $N$, then $E=\Eins$ and $N=M$. 
\vskip1.0mm
{\bf Borchers' Theorem} \cite[Thm.\ II.5.2]{B}: Let $M$ be a von
Neumann algebra on a Hilbert space $\HH$ and $\Omega\in \HH$ a vector
which is cyclic and separating for $M$. Let $U(s)$ be a unitary group
preserving $\Omega$ such that $U(s)MU(-s)\subset M$ for $s>0$. If the
generator of $U$ is positive, then one has
$$\Delta^{it}U(s)\Delta^{-it} = U(\hbox{e}^{-2\pi t}s).\eqno(A.4)$$ 
\vskip1.0mm
Choosing $M=B(\RR_+)$ in a M\"obius covariant chiral QFT, $\Omega$ the
vacuum, and $U(s)$ the translation subgroup of the M\"obius group, the
spectrum condition implies by Borchers' theorem that the modular group
has the same commutation relations with the translations as the
dilations (scaled by a factor of $-2\pi$). This implies \cite{GL,DLR}
that the modular group coincides with the unitary representers of the
scaled dilation subgroup up to a unitary cocycle $z(t)$ taking values
in the center of the gauge group of $B$
$$\Delta^{it}_{(B(I),\Omega)}=U(\Lambda_I(-2\pi t))\cdot z(t).\eqno(A.5)$$
The BW property is the statement that this cocycle is trivial. 

If $B$ is local, it has been shown previously \cite{GL} that the BW
property holds, by extending the theory to a net on the circle on
which the M\"obius group acts globally, and using the triviality of
$U(R(2\pi))$ where $R$ is the subgroup of rotations of the circle. 
This argument has been generalized \cite{DLR} to $\ZZ_2$-graded
(fermionic) theories, but it fails for more general non-local nets for
which $U(R(4\pi))\neq 1$. Our result (P2) provides an alternative
sufficient condition for the BW property of non-local chiral nets
which does not depend on the spectrum of the rotations.

\end{document}